\documentclass{article}
\usepackage{lajolla2006}
\usepackage{graphicx}
\usepackage{subfigure}
\usepackage{amssymb}

\newcommand{\vs}{\textit{vs.}}

\newcommand{\sqrtsNN}{\mbox{$\sqrt{s_{NN}}$}}
\newcommand{\jpsi}{\mbox{$J/\psi$}}
\newcommand{\gev}{\mbox{$\mathrm{GeV}$}}

\newcommand{\gevcc}{\mbox{$\mathrm{GeV/c^2}$}}
\newcommand{\dEdx}{\mbox{$dE/dx$}}

\frompage{000} \topage{000}                                              
\def\rightmark{Quarkonium Progress in STAR}
\def\leftmark{M. Calder\'{o}n de la Barca S\'{a}nchez et al.}

\title{Progress on Quarkonium Studies in STAR}
\authors{
{Manuel Calder\'{o}n de la Barca S\'{a}nchez $^1$ for the STAR Collaboration 
}\\[2.812mm]
{\normalsize
\hspace*{-8pt}$^1$ University of California, Department of Physics, \\
One Shields Ave, Davis CA 95616 USA\\[0.2ex]
}}

\abstract{We review recent progress on the study of quarkonium
production with the STAR detector.  We discuss the results of a
$\jpsi \rightarrow e^+e^-$ test trigger for $p+p$ collisions during
Run V at RHIC.  For the Au+Au data from Run IV, we discuss the
implementation of a test trigger for $\Upsilon \rightarrow e^+e^-$.
We obtain an upper limit on $\Upsilon$ production from the test
trigger data.  The prospects for future runs with upgraded detectors
are outlined.}


\begin{document}

\maketitle
\setcounter{page}{1}

\section{Introduction}\label{sec:intro}
In the study of relativistic heavy-ion collisions, one of the main
goals is to create a system of deconfined quarks and gluons in the
laboratory and to study its properties.  In the experiments
performed at the Relativistic Heavy Ion Collider (RHIC) the
expectation is that the state formed in the collision of the nuclei
will be the one described by high-temperature Quantum Chromodynamics
(QCD), the Quark-Gluon Plasma (QGP).  One of the key properties that
needs to be addressed is if the state produced requires a
description that invokes deconfined quarks and gluons. This question
is a longstanding one in the field.  One of the avenues which
experimentalists have used to explore the issue of deconfinement
dates back more than 20 years: the idea that the charmonium states
will be suppressed due to screening in the QGP \cite{Matsui:1986dk}.
More recently, studies in the Lattice QCD have aimed to put these
ideas on first-principles footing, and the latest avenues of inquiry
invoking studies of the charmonium correlators and the modification
of the spectral functions suggest that perhaps the suppression might
not be due to color screening\cite{Mocsy:2005qw}.  From the
experimental side, there are now measurements of \jpsi\ production
from SPS to RHIC energies.  The magnitude of the suppression
observed by the NA50 collaboration \cite{Cortese:2005ns} at
$\sqrtsNN=17.2\ \gev$ on Pb+Pb collisions, which has been
corroborated with much improved statistics by the NA60 experiment
(In+In at the same beam energy)\cite{Lourenco:2005zz}, is
intriguingly similar to the one from the PHENIX preliminary results
at $\sqrtsNN=200\ \gev$\cite{PereiraDaCosta:2005xz}. One possible
interpretation of the results is that the charmonium excited states
such as the $\psi'$ and $\chi_c$ could be suppressed while the
prompt $\jpsi$ might still survive at least up to RHIC energies.  It
has become increasingly clear that further progress can be achieved
by the measurement of a full spectroscopy of quarkonium states,
including both charm and beauty.  A measurement of the ratios of the
various quarkonia states in-medium might be a key to connecting with
first principles Lattice QCD above the deconfinement temperature
\cite{Digal:2001ue}.

In this paper, we will report on the progress in the STAR
collaboration towards the measurement of quarkonia.  We first
summarize the capabilities of the detector for the measurement of
electrons, since we will focus on the di-electron decay channel for
quarkonia.  We then discuss on the development of a di-electron
trigger, its expected performance, and compare with data taken with
a test implementation of the quarkonia triggers during RHIC Run IV.
We end by discussing the prospects for the implementation of the
triggers for the long RHIC p+p Run VI.

\section{Electron Identification in STAR}
\label{sec:ElectronID} The STAR experiment\cite{Ackermann:2002ad}
has as its main component a large-acceptance Time Projection Chamber
(TPC) surrounded by a solenoidal magnetic field, covering $|\eta|<1$
for tracks crossing the entire TPC volume. The TPC is complemented
by the addition of Electromagnetic Calorimetry (EMC), both as a
barrel\cite{Beddo:2002zx} surrounding the TPC and an
endcap\cite{Allgower:2002zy} in the forward region $1 < \eta < 2$ .
The EMC increases the electron identification capabilities and
allows to trigger on electromagnetic showers with energy depositions
of $\sim 1~\gev$ or above. For the triggers discussed in this paper,
we used the Barrel EMC (BEMC) only.  During Run IV, the BEMC
coverage with operating detectors was limited to $0 < \eta < 1 $.
For Run VI, the BEMC has been completed, affording the full
$|\eta|<1$ coverage.

The offline electron identification capabilities have been discussed
in Ref.~\cite{Suaide:2004rp}.  The main thrust is to combine the TPC
energy loss measurement (\dEdx) with the energy of the shower
measured in the BEMC.  The electrons will deposit all their energy
in the calorimeter, so the momentum $p$ measured in the TPC should
be equal to the energy $E$ measured in the calorimeter.  Hadrons
will not deposit all their energy, so a selection on the ratio
$E/p\sim1$ is a first step in the electron selection.  Next, from
the TPC $\dEdx\ \vs\ p$ measurement, a selection of tracks
consistent with the electron hypothesis is performed.  For example,
a selection at the $2\sigma$ level (after the BEMC selection),
yields an electron efficiency of 95\%, with 80\% purity. For this
paper, the more important aspect are the online capabilities, which
we now turn our focus to.

\section{Quarkonia Triggers}
\label{sec:Triggers}

At the trigger level, the information available EMC.  Since the EMC
is sensitive to both electrons and photons, a further photon-veto is
desirable for the $\jpsi$ trigger in order to reduce backgrounds.
For the $\Upsilon$ trigger, the larger mass allows to place a more
stringent requirement on the measured energies, keeping the event
rate low enough so no additional photon-veto was deemed necessary.
We discuss each trigger scheme now in greater detail.

\subsection{Level-0}
\label{sec:Level0} The STAR Level-0 trigger (the hardware trigger)
consists of a four layer tree structure of data storage and
manipulation boards (DSM). A trigger decision from any of the
trigger detectors is made every 104 ns, i.e., for each RHIC bunch
crossing. For the calorimeters, only a subset of the information is
available at the bunch crossing rate of 9.4 MHz. The BEMC towers are
combined in groups of 16 in their front-end electronics. For each
group, the energy sum and the largest signal within the group
(``High-tower'') is available with a resolution of 6 bits.

\begin{figure}[htb]
\vspace*{-.3cm}
\begin{center}
\includegraphics[width=0.45\textwidth]{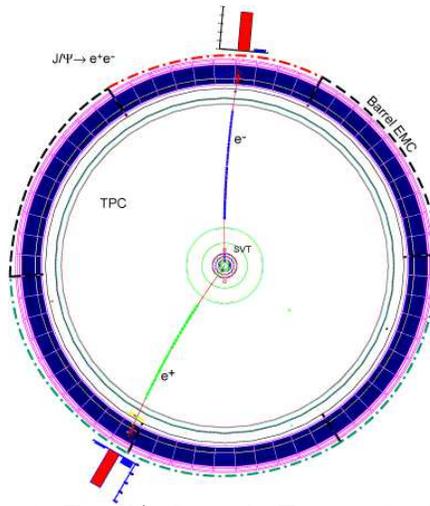}
\vspace*{-.5cm} \caption[]{The \jpsi\ Level-0 ``Topology'' trigger.}
\end{center}
\label{fig:jpsiL0}
\end{figure}

For the $\Upsilon$, the High-tower Level-0 trigger was found to give
the best combination of efficiency and background rejection. For the
\jpsi\, a more elaborate ``Topology'' trigger was implemented.  The
BEMC was divided into 6 separate sections in azimuth, each having
$20\times20$ towers (a coverage of roughly $1\times1$ in $\eta$ and
$\phi$).  The trigger then looked for High-towers in those sections
consistent with a back-to-back topology: if a High-tower was found
in one of the sections, the trigger logic looked for a High-tower in
any of the 3 opposite sections in $\phi$ to issue a trigger.  This
is illustrated in Fig.~\ref{fig:jpsiL0}.

The electron-positron pair from the \jpsi\ decay will create a
signal in BEMC towers roughly back-to-back.  In the example in the
Figure, one track heads towards the 12 o'clock section in the
detector and the logic looks for another High-tower in the 4, 6 or 8
o'clock sections (dot-dashed lines).  There is no requirement to
look at the adjacent 2 or 10 o'clock sections (dashed lines), an
adjacent section would be ignored by the trigger logic. This
effectively places a cutoff for $\jpsi$'s produced at large momenta,
which is $p_\perp= 5.4\ \gev/c$ at $y=0$. This is quite acceptable
since a $\jpsi$ with such transverse momentum will produce higher
energy electrons which can also be picked up by the standard single
High-tower trigger that runs at a higher energy threshold.

\subsection{Level-2}
\label{sec:Level2} The STAR Level-2 trigger (the software trigger)
consists of a fast CPU running software algorithms, with the
requirement to issue decisions at a rate of 1 KHz. All the trigger
information, in particular the individual tower data with full
12-bit resolution can be sent to the Level-2 CPU.  The online
calibration achieves a resolution of $17\%/\sqrt{E}$ (compared to
$14\%/\sqrt{E}$ for the final offline performance).

The Level-2 quarkonium trigger uses the individual tower information
to tag any tower above threshold as an electron candidate.  In order
to reject photons, another of the trigger detectors is used.  The
Central Trigger Barrel (CTB)\cite{Bieser:2002ah}, which consists of
scintillator slats covering the TPC and located in front of each
BEMC module, is used for this purpose. Charged particles will
generate signals in the scintillator, so a coincidence between the
BEMC tower and the slat that is in front of the tower is required by
the trigger algorithm.  Each slat covers 20 individual towers, so
the veto works best for low multiplicity p+p events.  In the
high-multiplicity environment of a Au+Au collision, the rejection
power of the $\jpsi$ trigger is sufficiently degraded that it is not
worthwhile to run a trigger.

To achieve a better energy resolution, a simple 3-tower cluster is
created instead of using the single tower energy. Since the charge
of the particle is not known, a straight line is assumed, whereby we
take the position vector of the cluster and make a straight line
from the cluster to the center of the detector. With the set of
candidate clusters, all pair combinations are made in order to
calculate an approximate invariant mass given by $m^2\simeq E_1 E_2
(1-\cos \theta)$, where $E_1$, $E_2$ are the energies of each
cluster in the pair, and $\theta$ is the 3-D opening angle between
the straight-line vectors.

Since the total time for all Level-2 algorithms to run must be kept
below 1 KHz, the aim was to keep the CPU time for each algorithm
that runs on Level-2 below $t\leq 300\ \mu s$.  The quarkonium
triggers were the only ones using Level-2 in Run IV so timing was
not of the utmost concern at that time.  After these triggers blazed
the trail and proved a working Level-2 system, there are now several
algorithms implemented and running in Run VI, so the timing
requirements are now a reality.

The two quarkonia triggers implemented in STAR can then be
summarized as follows.  For the $\jpsi$, the full trigger consisted
of the Topology trigger at Level-0 followed by the Level-2
invariant-mass algorithm with the photon veto. The trigger was
deployed in Run V only for p+p collisions. The Level-0 stage of the
trigger was first commissioned for a period of $\sim 4$ weeks,
followed by Level-2 commissioning until the end of the p+p run. For
the $\Upsilon$, the trigger path involved the High-tower Level-0
followed by the Level-2 invariant-mass algorithm (with no additional
photon veto). The $\Upsilon$ trigger can withstand the
high-multiplicity environment of a Au+Au collision, and was
commissioned during Au+Au data-taking in Run IV.

\section{Results}
\label{sec:Results}

\begin{figure}[htb]
\vspace*{-.5cm}
\subfigure{\includegraphics[width=0.5\textwidth]{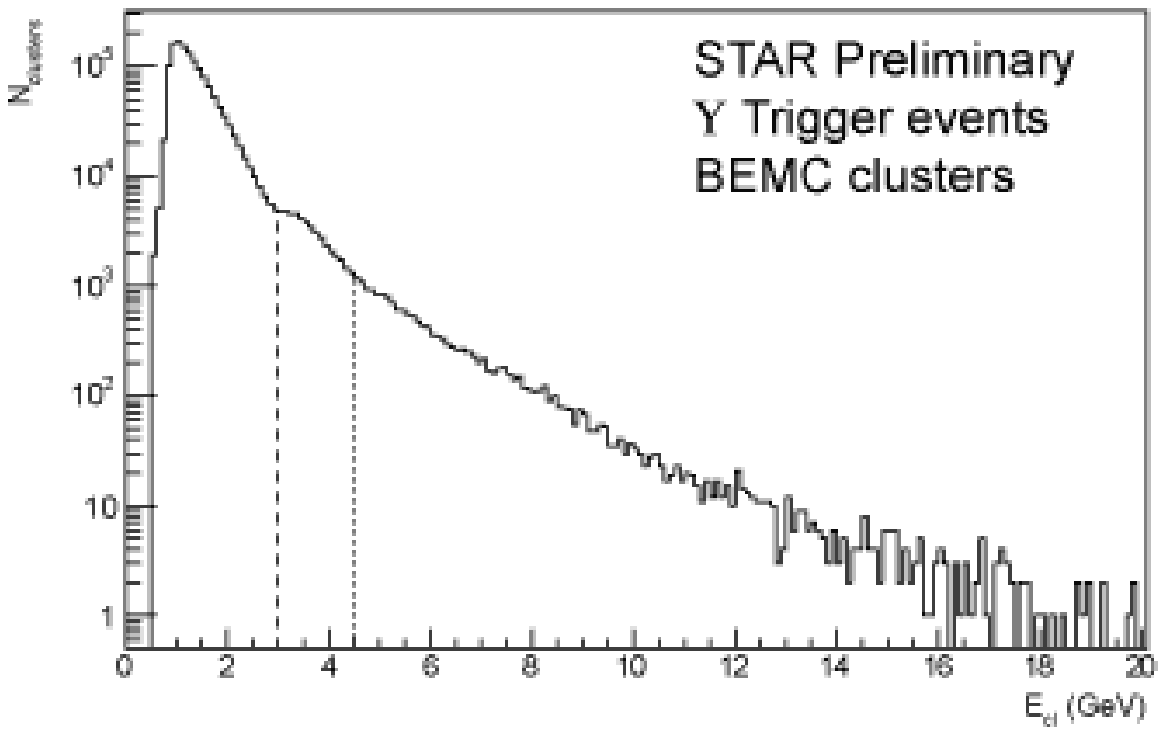}}
\subfigure{\includegraphics[width=0.5\textwidth]{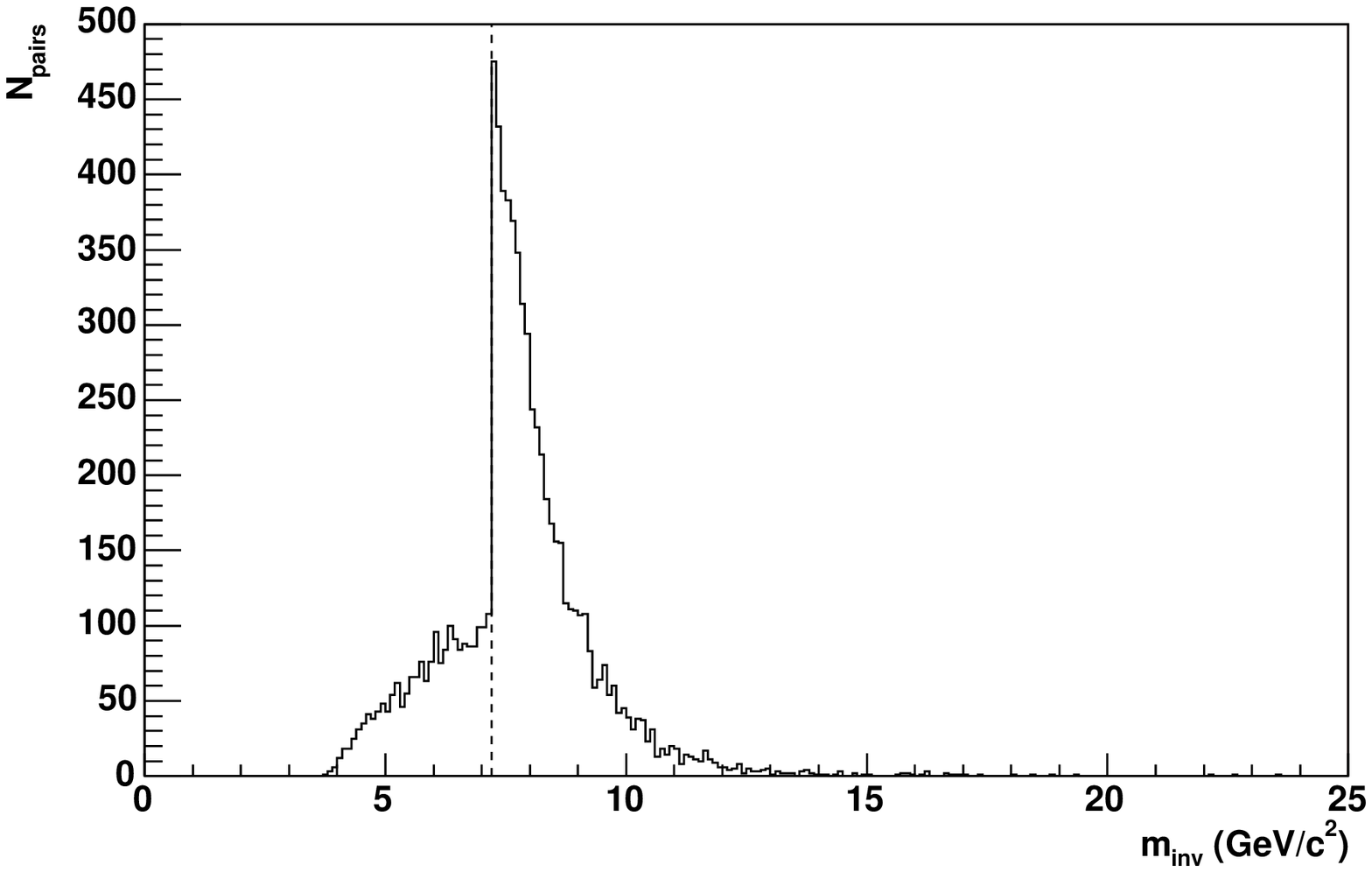}}
\vspace*{-1cm} \caption[]{Left Panel: The energy spectrum of the
$\Upsilon$ triggered events, where the High-tower threshold was at
$E_\perp\simeq3.5\ \gev$. Right Panel: The Level-2 pair invariant
mass for the $\Upsilon$ trigger sample.  The mass threshold for the
$\Upsilon$ Level-2 algorithm was placed at $m>7\ \gev$.}
\label{fig:clusterAndMass}
\end{figure}

We now discuss the performance of the quarkonia triggers in STAR.
From the $\Upsilon$ triggers, we can first scrutinize the output at
each of the trigger stages.  The cluster energies obtained after
Level-0 are shown in Fig.~\ref{fig:clusterAndMass}, left panel.

The calibration of the BEMC was expected to be uniform in transverse
energy, $E_\perp$.  There was an unfortunate miscalibration which
was not discovered until after the run was over.  The figure shows
the cluster energies for the triggered sample.  The $E_\perp$
threshold was set at 3.5 \gev.  Since for the algorithm we use $E$,
this is what appears in the ordinate of the Figure. The conversion
from $E_\perp$ to $E$ should produce some smearing, but there was an
additional smearing due to the miscalibration.  This has been
corrected for future runs.  The right panel of
Fig.~\ref{fig:clusterAndMass} shows the invariant mass of the
candidate pairs seen at Level-2.  The threshold at $m>7\ \gev$ is
clearly visible.  Similar results are obtained for the $\jpsi$
triggered events with the corresponding threshold ($m\gtrsim2\
\gev$).

The $\Upsilon$ trigger data from Au+Au collisions in Run IV was then
analyzed offline\cite{Kollegger:2006}.
\begin{figure}[htb]
\vspace*{-.5cm}
\begin{center}
\subfigure{\includegraphics[width=0.5\textwidth]{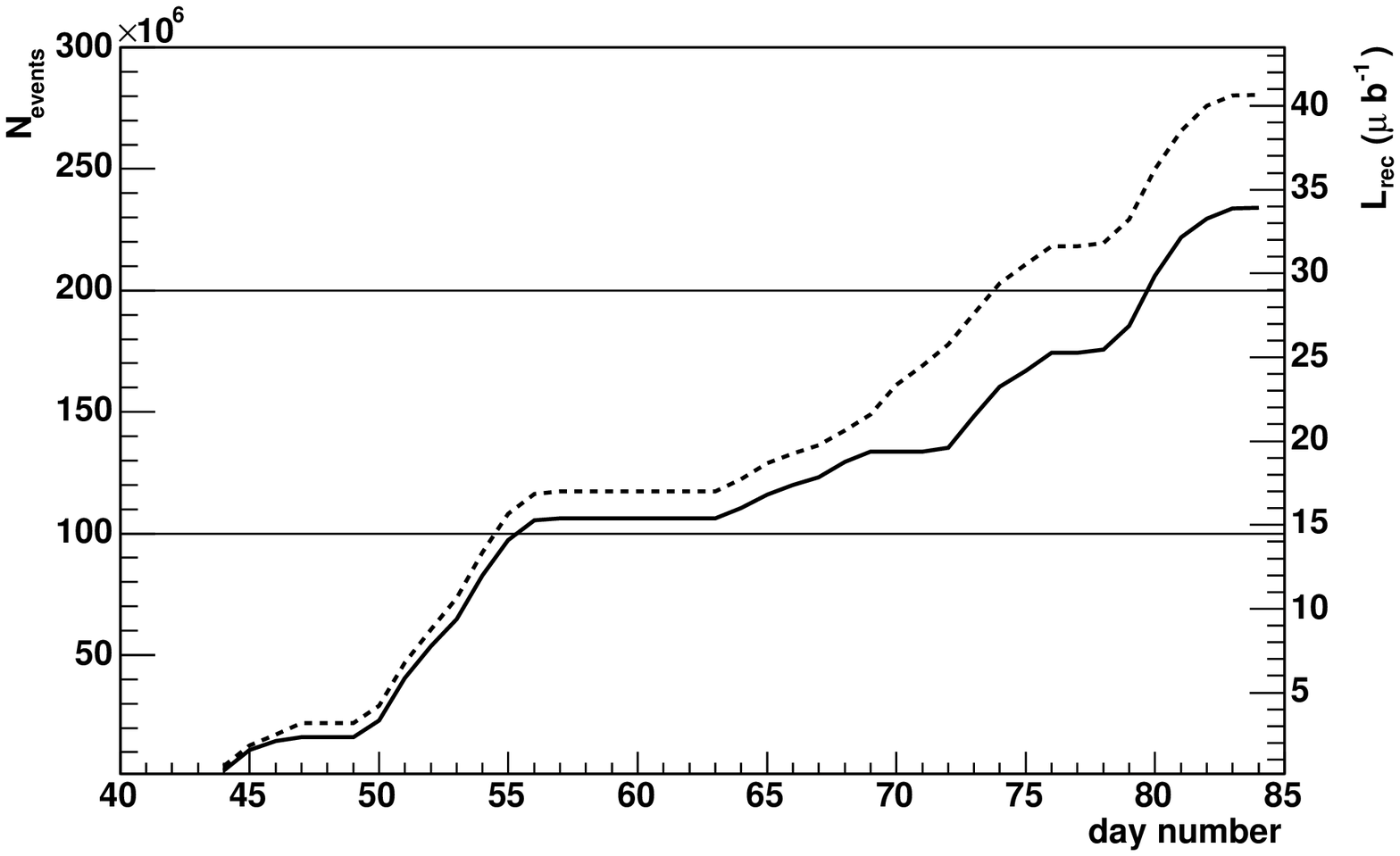}}
\subfigure{\includegraphics[width=0.35\textwidth]{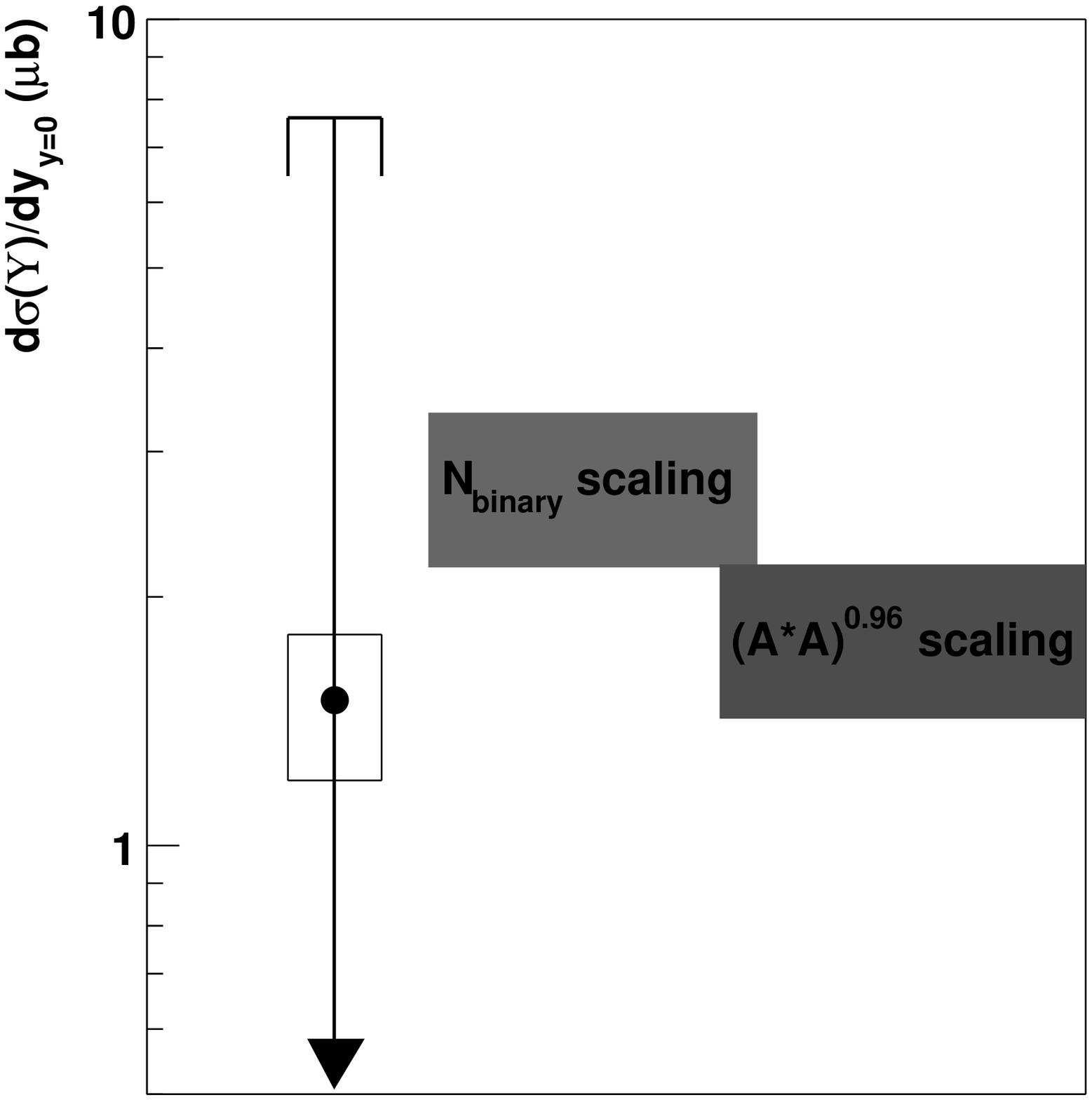}}
\vspace*{-.2cm} \caption[]{Left Panel: The integrated luminosity for
Au+Au minbias events seen at Level-0.  The dashed line is the total
and the full line is the part where the $\Upsilon$ trigger was live.
Right Panel: Upper limit for $\Upsilon$ production from the null
result from the Au+Au sample collected by STAR in Run IV.}
\end{center}
\label{fig:upsilonResults}
\end{figure}
The right panel of Fig.~\ref{fig:upsilonResults} shows the
integrated luminosity for Run IV Au+Au running.  The full line is
the minimum bias luminosity seen at Level-0 and the dashed line is
the fraction when the $\Upsilon$ trigger was live.  The ordinate
axis is the day number (in 2004). (The period around day 60 where no
events are counted is from a dataset where the STAR magnetic field
was ramped down from its full magnitude of 0.5 T to 0.25 T as part
of a $D^{*+}$ search.)  The search for a signal in the $\Upsilon$
data sample yielded a null result.  With the integrated luminosity
achieved during the run, we estimate an upper limit for the
$\Upsilon$ cross section, shown in the right panel of
Fig.~\ref{fig:upsilonResults}.  The upper limit is consistent with
binary scaling of the $\Upsilon$ cross section from p+p to Au+Au, as
well as with scaling with $(AA)^{\alpha}$ where $\alpha=0.96$
obtained from E866 results \cite{Leitch:1999ea}. The addition of
BEMC coverage in the $-1<\eta<0$ region, together with the correct
calibration of the BEMC, should both allow for a measurement of a
$4\sigma$ signal or better in a future Au+Au run.

\begin{figure}[htb]
\vspace*{-.3cm}
\begin{center}
\subfigure{\includegraphics[width=0.4\textwidth]{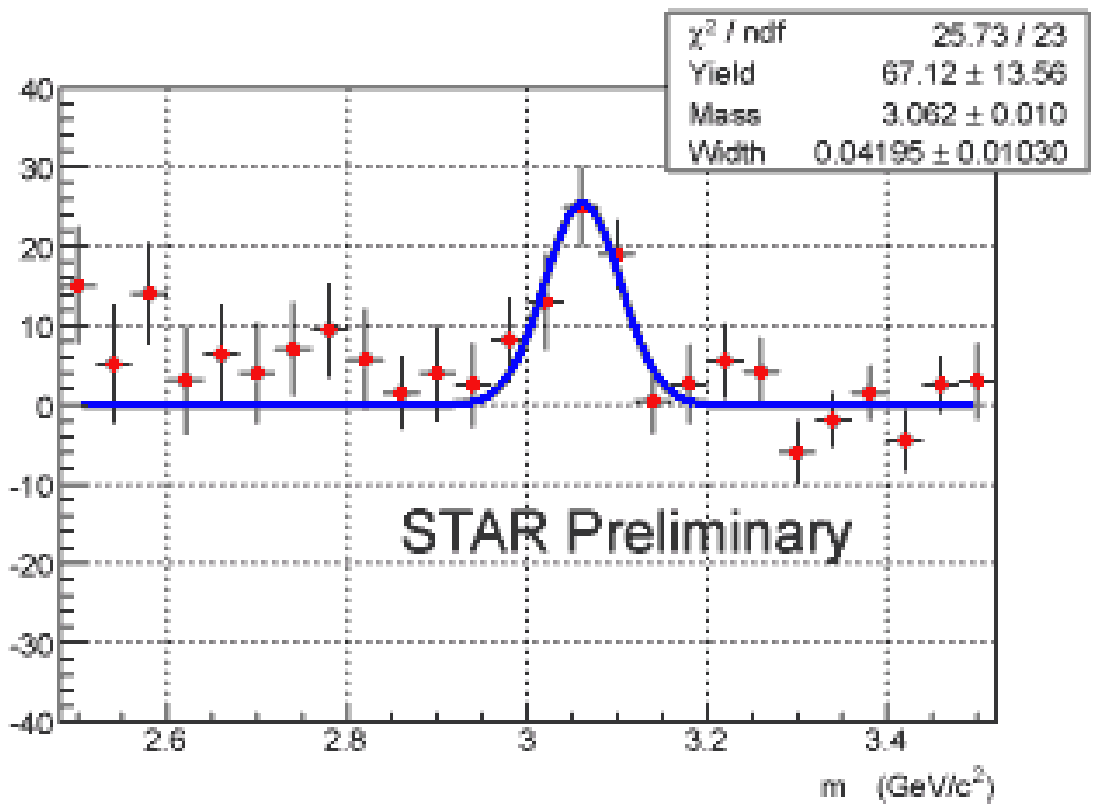}}
\subfigure{\includegraphics[width=0.4\textwidth]{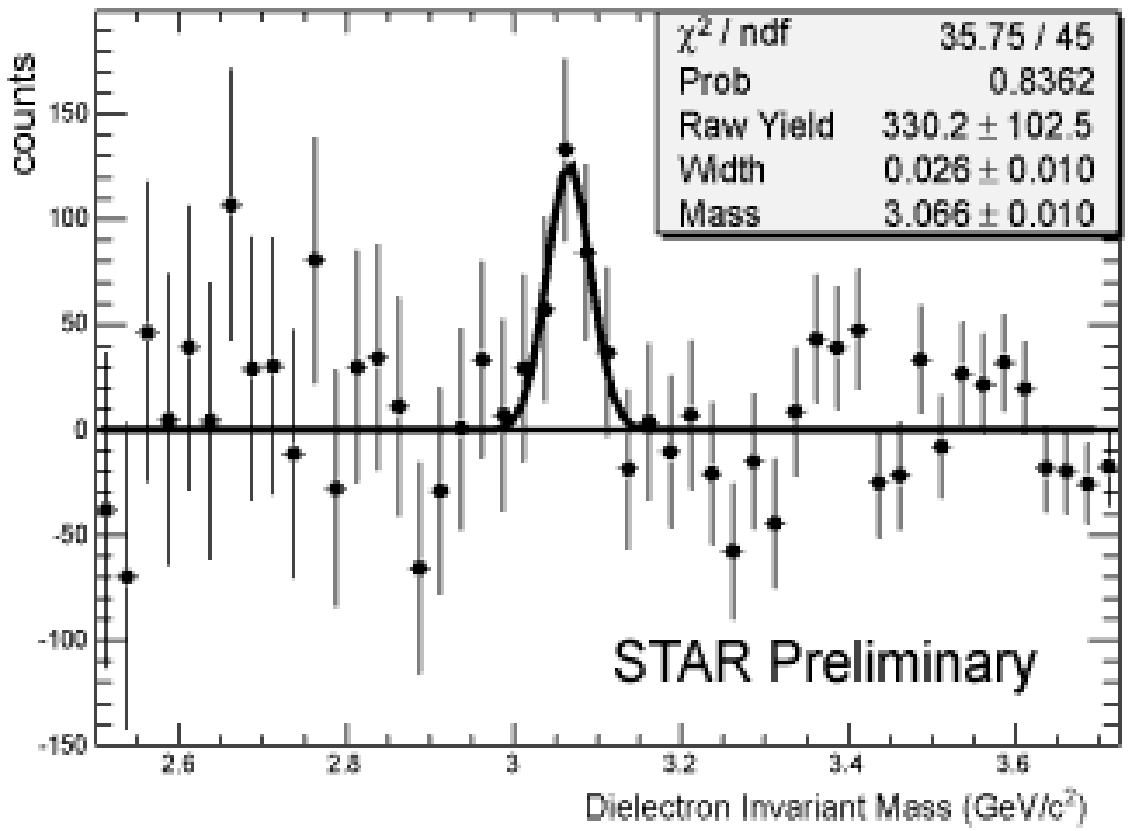}}
\vspace*{-.6cm} \caption[]{Left Panel: Raw $\jpsi$ signal in p+p
obtained from the commissioning of the $\jpsi$ trigger during Run V.
Right Panel: Raw $\jpsi$ signal from the offline analysis of Au+Au
data from Run IV.}
\end{center}
\label{fig:jpsiResults}
\end{figure}

The $\jpsi$ analysis of the trigger data from the commissioning
period with p+p collisions is shown in Fig.~\ref{fig:jpsiResults},
left panel.
For comparison, the signal obtained from the analysis of the Au+Au
data collected in Run IV is also shown \cite{Gonzalez:2006}.  The
$\jpsi$ Level-0 trigger achieved a rejection power, defined as the
number of background events that get rejected for every one that is
triggered, of $\simeq~10^3$.  The addition of the Level-2 algorithm
increased the rejection by a factor of $\sim6$.  The raw yield
obtained in the trigger sample was consistent with simulations of
the cross section folded with the detector acceptance and the
trigger efficiency ($\epsilon=0.36$). The mass resolution obtained
offline for the triggered data-set is found to be $0.042\ \gevcc$,
in excellent agreement with simulations.  The mass resolution
improves slightly in the Au+Au environment due to an improved vertex
reconstruction with the larger track multiplicity.  The
reconstructed peak position is lower than the PDG value\ due to
electron bremsstrahlung in the detector. The simulation results
incorporating this effect gave a mass of $3.08\ \gevcc$, slightly
above the $3.06\ \gevcc$ by $\sim2\sigma$.  The performance of the
trigger is in general consistent with our expectations.

\section{Future Prospects}
\label{sec:Future}

The quarkonia triggers in STAR have been successfully implemented
and the data from the test runs has proved that they are performing
as expected.  The Level-2 framework has proven to be a very useful
tool, and is currently heavily used in Run VI for searches of jets
and di-jet events, for example.  As part of the p+p program for Run
VI, STAR expects to collect enough statistics for a 3-4$\sigma$
signal in the $\jpsi\rightarrow e^+e^-$ channel.

\begin{figure}[htb]
\vspace*{-.2cm}
\begin{center}
\includegraphics[width=0.45\textwidth]{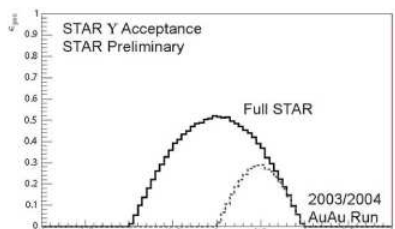}
\vspace*{-.7cm} \caption[]{The acceptance as a function of rapidity
for $\Upsilon\rightarrow e^+e^-$ for the STAR BEMC in Run IV (dashed
line) and in Run VI (full line)}
\end{center}
\label{fig:acceptance}
\end{figure}

The quarkonia triggers will benefit from the full installation of
the Barrel EMC (completed for Run VI), which provides coverage in
the region $|\eta|<1$ over full azimuth.  Thanks to this increased
coverage, the acceptance for the di-electron events increases by a
factor 4 over the acceptance from Run IV, as shown in
Fig.~\ref{fig:acceptance}.

Longer term upgrades are also being considered.  One avenue of
improvement is to allow for a knowledge of the vertex position at
Level-0.  By upgrading the hardware and the electronics of the
Time-of-Flight (TOF) Vertex Position Detector (used as a start-time
for the TOF) a vertex resolution of $\sim 1\ cm$ could be available
at Level-0. This would allow to select events near the center of the
STAR detector; a benefit for all STAR analyses.  For the quarkonia
triggers, the knowledge of the vertex position would also allow for
an improved online mass resolution.

Additional capabilities are also brought about by the full-barrel
TOF upgrade (expected in 2009).  The full-barrel TOF detector will
enhance the electron identification capabilities over the same
acceptance as the TPC and BEMC.  At the trigger level, the finer
segmentation of the TOF will also be an improvement over the current
scheme of using the CTB as a photon-veto.  The program for the
future runs in STAR offers the exciting opportunity for studies in
quarkonium production.

\section{Conclusions}
\label{sec:Conclusions}

The quarkonia studies in heavy-ion collisions are a means to study
the question of deconfinement in a QGP.  This necessitates a full
program of study of charmonium and bottomonium states.  The STAR
experiment has implemented two dedicated calorimeter-based triggers
for the study of charmonium in p+p collisions and of bottomonium up
to the highest-multiplicity Au+Au collisions. These have been put
into production for p+p collisions in RHIC Run VI, where the barrel
EMC is fully installed and instrumented.  From the data collected
during Run VI, we expect to make a 3-4$\sigma$ measurement of
$\jpsi$ production in p+p at $\sqrtsNN=200\ \gev$.

%
%
%
%

\vfill\eject
\end{document}